\documentclass[aps,prd,reprint,onecolumn,notitlepage,superscriptaddress]{revtex4-1}

\usepackage{lipsum}
\usepackage{amsmath}
\usepackage{graphicx}
\usepackage{color}
\usepackage{tikz}
\usepackage{verbatim}

\usepackage{amssymb}
\usepackage{bm}

\newcommand{\rom}[1]{\textup{\uppercase\expandafter{\romannumeral#1}}}

\begin{document}

\title{Geodetic Motion Around Rotating Blackhole in Nonlocal Gravity}
\author{Utkarsh Kumar\footnote{email:kumaru@ariel.ac.il},
       }
\affiliation{Department of Physics, Ariel University, Ariel 40700, Israel}

\author{Sukanta Panda\footnote{email:sukanta@iiserb.ac.in},
        Avani Patel\footnote{email: avani@iiserb.ac.in}}
        \affiliation{ Indian Institute of Science Education and Research Bhopal,\\ Bhopal 462066, India}



\begin{abstract}
Recently the non-local gravity theory has come out to be a good candidate for an effective field theory of quantum gravity and also it can provide rich phenomenology to understand late-time accelerating expansion of the universe. For any valid theory of gravity, it has to surmount solar system tests as well as strong field tests. Having motivations to prepare the framework for the strong field test of the modified gravity using Extreme Mass Ratio Inspirals(EMRIs), here we try to obtain the metric for Kerr-like blackhole for a non-local gravity model known as RR model and calculate the shift in orbital frequencies of a test particle moving around the blackhole. We also derive the metric for a rotating object in the weak gravity regime for the same model. 

\end{abstract}

\maketitle
\section{INTRODUCTION}
Nonlocal gravity theories have recently gained attention because of its ability to explain the late time cosmology unified with inflationary era. Especially so-called $RR$ model proposed by \cite{Maggiore:2014sia} has been shown to explain CMB+BAO+SNIa+RSD data as well as $\Lambda$CDM\cite{Amendola:2019fhc}.  Let us first write the action for the $RR$ model :
\begin{eqnarray}
S = \frac{1}{2\kappa^{2}}\int d^{4}x \sqrt{-g}\Big[ R + \frac{\mu^{2}}{3} R\frac{1}{\Box^2}R\Big] + \mathcal{L}_{m}.  \label{action}
\end{eqnarray}
For this model we require a additional mass term, $\mu$, that gives a constraint $ \mu = 0.283 H_{0}$, where $H_{0}$  is Hubble parameter today, to give the viable cosmology.
Equation of motion corresponding to action~\eqref{action} is
\begin{equation}
\kappa^{2} T_{\alpha \beta}  =  G_{\alpha \beta} - \frac{\mu^{2}}{3}\Big\{ 2\Big(G_{\alpha
 \beta} - \nabla_{\alpha}\nabla_{\beta} + g_{\alpha \beta} \Box \Big)S  + g_{\alpha \beta} \nabla^{\gamma}U \nabla_{\gamma}S   - \nabla_{(\alpha}U \nabla_{\beta)}S  -\frac{1}{2}g_{\alpha \beta} U^{2} \Big\} , \label{gumu}
\end{equation}
with $ U = - \frac{1}{\Box}R$  and $ S = - \frac{1}{\Box}U$, where, $G_{\alpha \beta}$ is the Einstein tensor and $T_{\alpha \beta}$ is the energy-momentum tensor of the matter. The gravitational waves propagation in $RR$ model has been studied in \cite{Belgacem:2018lbp}.

The orbits traversed by the small compact object(SCO) of mass $m$ orbiting around a super massive blackhole(SMBH) of mass $M$ is generally known as extreme mass ratio inspirals(EMRIs). Since the mass ratio $M/m$ of two objects is $\sim 10^4-10^8M_{\odot}$, the  motion of the SCO around the SMBH can be approximated as trajectory of a point particle along the geodesics of the SMBH spacetime. The structure of the spacetime can be reflected in orbital frequencies of the point particle whose imprints finally can be seen in gravitational waves emitted by SCO. Here, we aim to calculate the shift in orbital frequencies of geodesic motion of the test particle around rotating blackhole due to nonlocal correction in RR model following the treatment prescribed in \cite{Vigeland:2009pr}.


\section{Rotating Object in the Weak Gravity Regime in RR Model}
We consider the linearized gravity limit of the field equation written in Eq.~\eqref{gumu}. In linearized gravity limit we take $g_{\alpha\beta} $ as $g_{\alpha\beta} = \eta_{\alpha\beta} + h_{\alpha\beta},\;\;\;|h|\ll 1$, where $h(\alpha\beta)$ is a small perturbation around Minkowski background $ \eta_{\alpha\beta}$. Under this approximation, one can find the expressions for Riemann Tensor, Ricci Tensor and Ricci scalar as follows
\begin{eqnarray}
 R_{\gamma\alpha\delta\beta} &=& \frac{1}{2} \Big( \partial_{\delta}\partial_{\alpha} h_{\gamma\beta} + \partial_{\beta}\partial_{\gamma} h_{\alpha\delta} -\partial_{\beta}\partial_{\alpha} h_{\gamma\delta} -\partial_{\delta}\partial_{\gamma} h_{\beta\alpha}\Big) , \label{weakrt} \\
R_{\alpha\beta}& =& \frac{1}{2}\Big( \partial^{\gamma}\partial_{\alpha}h_{\gamma \beta} + \partial_{\beta}\partial_{\gamma} h_{\alpha}^{\gamma} -\partial_{\beta}\partial_{\alpha}h -\Box h_{\alpha\beta} \Big) ,  \label{weakri} \\ 
R &= &\partial_{\alpha}\partial_{\beta}h^{\alpha\beta} - \Box h . \label{weakrs}
\end{eqnarray}
Then the field equation \eqref{gumu} becomes
\begin{equation}
\begin{split}
2\kappa^2 T_{\alpha\beta} &= -\Bigg[\Box h_{\alpha\beta} -\partial_{\gamma} \partial_{(\alpha}h^{\gamma}_{\beta)} + \Big(1  - \frac{2M^{2}}{3} \Box^{-1}  \Big)(\partial_{\alpha}\partial_{\beta}h + \eta_{\alpha\beta}\partial_{\gamma}\partial_{\delta}h^{\gamma\delta}) \\&-\Big( -1  + \frac{2M^2}{3} \Box^{-1}  \Big) \eta_{\alpha\beta} \Box h  + \frac{2M^{2}}{3} \Box^{-2}\nabla_{\alpha}\nabla_{\beta}\partial_{\gamma}\partial_{\delta}h^{\gamma\delta}\Bigg].
\end{split}    \label{rearrgmunu}
\end{equation}

\subsection{Spacetime Solution around Rotating Object in Linearized Gravity Limit} 
Starting with a generic spherically symmetric and static metric
\begin{equation}
ds^{2} = -(1 + 2\Phi ) dt^2 + 2 \vec{h}.d \mathtt{x} dt + (1 - 2\Psi)d \mathtt{x}^2,
\label{metric1}
\end{equation}
and the stress-energy tensor for the rotating object having energy density $\rho=M\delta^3(\vec{r})$ with mass $M$ and angular velocity $v_i$ given by $T_{00} = \rho, \,\, T_{0i}  =  -\rho v_{i},$ we can solve the field equation in \eqref{gumu} and convert the solution into Boyer-Lindquist coordinates $(t, r, \theta, \phi)$ to obtain the rotating metric as 
\begin{equation}
ds^{2}   =  -(1 + 2\Phi)dt^{2} +4\frac{J\sin^{2}\theta}{M}(\Phi + \Psi) d\phi dt + (1- 2\Psi)( dr^2 + r^{2}d\theta^{2} + r^{2}\sin^{2}\theta d\phi^{2}),  \label{rotmet}
\end{equation}
where J is angular momentum, defined as $ v = \frac{r \times J }{M r^{2}}$ and $\Phi$ and $\Psi$ are given by\cite{Kumar:2018pkb}
\begin{equation}
\Phi(r) = \frac{GM}{r}\Big(\frac{e^{-\mu r}-4}{3}\Big),\;\;\;
\Psi(r)  =   \frac{GM}{r}\Big(\frac{-e^{-\mu r} - 2}{3}\Big), \label{phipsipot}. 
\end{equation}

\section{Geodetic Motion Around Rotating Blackhole in Nonlocal Gravity}
The metric for the spacetime around rotating blackhole in RR model was obtained by applying Demiański-Janis-Newman algorithm\cite{Newman:1965tw,Demianski:1972uza} on the spherically symmetric static solution\cite{Maggiore:2014sia} of the RR model in \cite{Kumar:2019uwi} as (in the form of $g_{\alpha\beta}=g_{\alpha\beta}^{Kerr}+b_{\alpha\beta}$)
\begin{eqnarray}
ds^{2}=  \left[ -1 + \frac{2GMr}{\Sigma}\Big( 1 + \frac{1}{6} \mu^{2}\Sigma \Big)\right]\; dt^{2} - 2a\sin^{2}\theta\left[ \frac{2GMr}{\Sigma}\Big( 1 + \frac{1}{6} \mu^{2}\Sigma \Big)\right]\;dt\;d\phi + \dfrac{\Sigma}{\Delta}\;dr^{2} + \Sigma \;d\theta^{2} + {\rm sin}^{2}\theta\left[\Sigma +\right.\nonumber\\
\left.\Big(1+\frac{2GMr}{\Sigma}\Big( 1 + \frac{1}{6} \mu^{2}\Sigma \Big)\Big)a^2 \rm sin^{2}\theta \right]\;d\phi^{2}.
\end{eqnarray}
where, $\Sigma  =  r^{2} + a^{2}\cos ^{2}\theta$ and $\Delta = \Sigma - 2GMr\Big( 1 + \frac{1}{6} \mu^{2}\Sigma \Big) + a^{2}\sin^{2}\theta$.

\begin{figure}
  \centering
   $
   \begin{array}{c c}
   \includegraphics[width=0.40\textwidth]{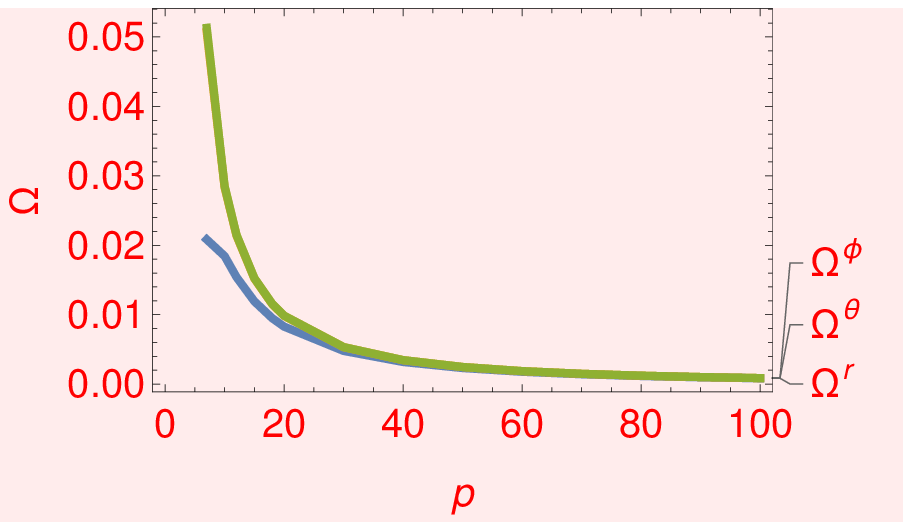} &
   \includegraphics[width=0.40\textwidth]{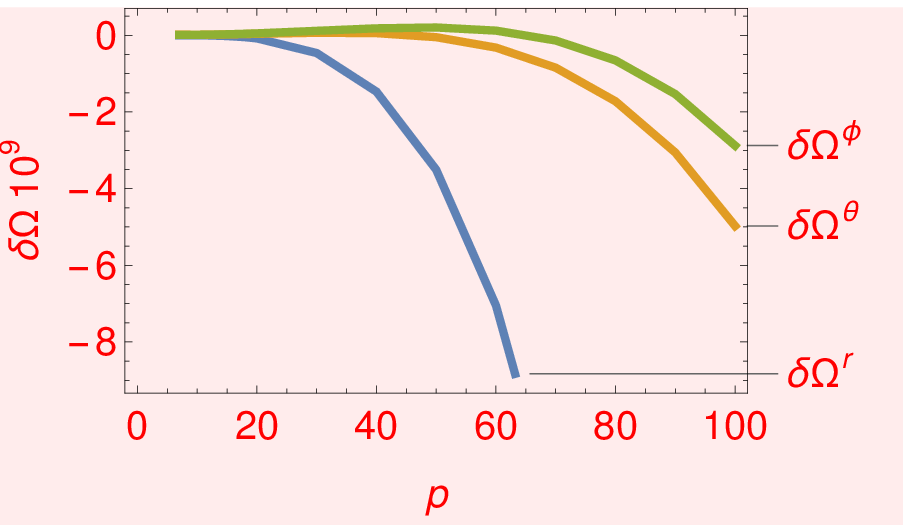} 
   \end{array}
   $ 
  \caption{$\Omega^i(=\omega^i/\omega^{t})$ Vs. p. Left : Observable orbital frequencies of the orbits in Kerr spacetime Right : shift in observable orbital frequencies due to nonlocal correction }
  \label{fig:orbitalfreq}
\end{figure}

Since the Kerr metric is independent of $t$ and $\phi$ it has two apparent symmetry and possesses two constants of motion i.e. energy $E$ as measured by observer at spatial infinity and axial component of angular momentum $L_z$. The equation of motion of a point mass $m$ moving along the geodesics of the Kerr metric is given by geodesic equation $\dot{u^{\mu}} + \Gamma_{\alpha\beta}^{\mu}u^{\alpha}u^{\beta}=0$, where $u^{\mu}$ is the four-velocity of the point mass and overdot denotes the derivative w.r.t. proper time $\tau$. Since the nonlocal correction term added to Einstein-Hilbert(EH) action is very small compared to EH term we can use canonical perturbation theory to solve the geodesic equations of nearly Kerr-like blackhole of nonlocal gravity model. A relativistic version of Hamilton-Jacobi method for the motion of a test mass in Kerr spacetime as proposed by Carter\cite{Carter:1968rr} shows that the motion is separable and it is attributed by three constants of motion : the Carter constant $Q$ in addition with $E$ and $L_z$. A further scheme of relativistic action-angle formalism to calculate the fundamental frequencies of the orbital motion in the Kerr geometry was provided by Schmidt\cite{Schmidt:2002qk}. According to canonical perturbation theory, if $\hat{\omega}^r,\hat{\omega}^{\theta}$ and $\hat{\omega}^{\phi}$ are orbital frequencies for Kerr blackhole then the orbital frequencies for the Kerr-like BH in nonlocal gravity are given by $\omega^i=\hat{\omega}^i+\delta\omega^i$, where $m\delta\omega^i=\partial\langle H_1\rangle /\partial\hat{J}_i$\cite{Vigeland:2009pr} and $\hat{J}_i$ is the action variable for the Kerr spacetime. The Hamiltonian of the system (test mass and SMBH) is given by $H = H_{Kerr} + H_1$, where $H_1 = -(m^2/2)b_{\alpha\beta}(dx^{\alpha}/d\tau)(dx^{\beta}/d\tau)$. $\langle H_1\rangle$ shows the averaged Hamiltonian $H_1$ over a period of the orbit in background spacetime.
We numerically calculate the orbital frequencies for Kerr spacetime(shown in left plot of Fig.[\ref{fig:orbitalfreq}]) and shift in frequencies from the Kerr frequencies due to nonlocal correction of $\mu^2R\frac{1}{\Box^2}R$(shown in right plot of Fig.[\ref{fig:orbitalfreq}]). Equivalent to set of $(E, L_z, Q)$, the orbital motion can be parametrized by set of parameters $(p, e, \theta_{min})$, where $p$, $e$ and $\theta_{min}$ are respectively semilatus rectum, eccentricity and turning point of $\theta$-motion of the orbit. The detailed discussion on calculation of shift in orbital frequencies and solution of geodesic equations of the spacetime which is slightly different from the Kerr spacetime will be done in our future work.
     
     
\section{CONCLUSION}
We calculate the axially symmetric stationary metric around the rotating object in RR model of nonlocal gravity by solving field equations in linearized gravity limit for axial symmetry. Using canonical perturbation theory we calculate the orbital frequencies of a test particle moving along the geodesic of the metric around rotating blackhole in $RR$ model of nonlocal gravity.
\section{ACKNOWLEDGEMENT}
This work was partially funded by DST (Govt. of India), Grant No. SERB/PHY/2017041. This work was presented at XXIII DAE-BRNS High Energy Physics Symposium 2018 held at IIT Madras, India.
 \bibliography{paper}

\end{document}